\documentclass{amsart}
\usepackage{times,amssymb,amscd}
\numberwithin{equation}{section}

\newtheorem{prop}{Proposition}
\newtheorem{theorem}[prop]{Theorem}

\theoremstyle{definition}

\numberwithin{prop}{section}
\newcommand{\m}{{\bf m}}

\newcommand{\n}{{\bf n}}

\newcommand{\uu}{{\bf u}}
\newcommand{\vv}{{\bf v}}
\newcommand{\e}{{\bf e}}
\newcommand{\I}{\mathcal{I}} 
\newcommand{\p}{{\bf p}}
\newcommand{\lb}{\overline{\ell}}
\newcommand{\yb}{\overline{y}}
\newcommand{\R}{\mathbb{R}}
\newcommand{\Z}{\mathbb{Z}}

\begin{document}

\title{Non-unitary minimal models, Bailey's Lemma and $N=1,2$ Superconformal 
algebras}

\author[L.~Deka]{Lipika Deka}
\address{Department of Mathematics\\
University of California\\
One Shields Ave\\
Davis, CA 95616-8633 U.S.A.}
\email{deka@math.ucdavis.edu}
\urladdr{http://www.math.ucdavis.edu/\~{}deka}

\author[A.~Schilling]{Anne Schilling}
\email{anne@math.ucdavis.edu}
\urladdr{http://www.math.ucdavis.edu/\~{}anne}
\thanks{\textit{Date:} December 2004}
\thanks{Supported in part by NSF grant DMS-0200774.}

\begin{abstract}
Using the Bailey flow construction, we derive character identities for the $N=1$ 
superconformal models $SM(p',2p+p')$ and $SM(p',3p'-2p)$, and the $N=2$ superconformal 
model with central charge $c=3(1-\frac{2p}{p'})$ from the nonunitary minimal models 
$M(p,p')$. A new Ramond sector character formula for representations of $N=2$ 
superconformal algebras with central element $c=3(1-\frac{2p}{p'})$ is given. 
\end{abstract}

\maketitle

\section{Introduction}
\noindent
Bailey's lemma is a powerful method to prove $q$-series identities of the 
Rogers--Ramanujan-type~\cite{B:1949}. One of the key features of Bailey's lemma
is its iterative structure which was first observed by Andrews~\cite{A:1984}
(see also~\cite{Paule:1985}). This iterative structure called the Bailey chain 
makes it possible to start with one seed identity and derive an infinite family 
of identities from it. The Bailey chain has been generalized to the Bailey 
lattice~\cite{AAB:1987} which yields a whole tree of identities from a single seed.

The relevance of the Andrews--Bailey construction to physics was first revealed
in the papers by Foda and Quano~\cite{FQ:1995,FQ:1996} in which they derived
identities for the Virasoro characters using Bailey's lemma.
By the application of Bailey's lemma to polynomial versions of the character
identity of one conformal field theory, one obtains character identities of
another conformal field theory. This relation between the two conformal
field theories is called Bailey flow. In~\cite{BMS:1995} it was demonstrated 
that there is a Bailey flow from the minimal models $M(p-1,p)$ to
$N=1$ and $N=2$ superconformal models. More precisely, it was shown that there
is a Bailey flow from $M(p-1,p)$ to $M(p,p+1)$, and from $M(p-1,p)$ to the $N=1$ 
superconformal model $SM(p,p+2)$ and the unitary $N=2$ superconformal model with 
central charge $c=3(1-\frac{2}{p})$. In the conclusions of~\cite{BMS:1995} it was
mentioned that this construction can also be carried out for the nonunitary
minimal models $M(p,p')$ where $p$ and $p'$ are relatively prime.
In this paper we consider the nonunitary case. We show that starting 
with character identities for the nonunitary minimal model $M(p,p')$ 
of~\cite{BMS:1997,W:2002}, characters of the $N=1$ superconformal models $SM(p',2p+p')$, 
$SM(p',3p'-2p)$ and of the $N=2$ superconformal model with central element 
$c=3(1-\frac{2p}{p'})$ can be obtained via the Bailey flow. 
We also give a new Ramond sector character formula for a representation of the 
$N=2$ superconformal model with central element $c=3(1-\frac{2p}{p'})$. 

The character identities obtained from the Bailey flow construction
are of Bose-Fermi type. The bosonic side is associated with the construction
of singular vectors of the underlying conformal field theory. The fermionic 
side is usually manifestly positive and reflects the quasiparticle structure
of the model.

The paper is organized as follows. In section~\ref{sec:bailey} we provide the 
necessary background about Bailey pairs and fermionic formulas of the $M(p,p')$
models. This section is added to make this paper self-contained. 
For details the reader should consult~\cite{BMS:1995,BMS:1997,BMSW:1997}. 
In section~\ref{sec:N1} the characters of the $N=1$ supersymmetric models 
$SM(2p+p',p')$ and $SM(3p'-2p,p')$ are derived using the Bailey flow. 
Explicit fermionic expressions for these characters are given. 
In section~\ref{sec:N2} the background regarding $N=2$ superconformal models is stated
and a new character for the Ramond sector is derived. Then it is demonstrated how to 
obtain the characters of the $N=2$ superconformal model with central element 
$c=3(1-\frac{2p}{p'})$ via the Bailey flow along with the explicit fermionic expressions
for these characters. In section~\ref{sec:conclusion} we conclude with some remarks.

\subsection*{Acknowledgment}
Special thanks to Professor Gaberdiel and Hanno Klemm for their help through the jungle
of literature regarding $N=2$ character formulas and their help regarding the spectral 
flow of $N=2$ superconformal algebras. 
We are grateful to both of them for their e-mail correspondences. 
We would also like to thank Professor Dobrev for helpful discussions.

\section{Bailey's lemma} \label{sec:bailey}

In this section we summarize Bailey's original lemma~\cite{A:1984,B:1949} and
the Bose-Fermi identities for the $M(p,p')$ minimal 
models~\cite{BM:1994,BMS:1997,FW:2001,W:2002}.

\subsection{Bilateral Bailey lemma}
A pair $(\alpha_n,\beta_n)$ of sequences $\{\alpha_n\}_{n\ge 0}$ and 
$\{\beta_n\}_{n\ge 0}$ is called a \textbf{Bailey pair} with respect to $a$ if 
\begin{equation}
\beta_n=\sum_{j=0}^n \frac{\alpha_j}{(q)_{n-j}(aq)_{n+j}}
\end{equation}
where 
\begin{equation*}
\begin{split}
(a)_n:=(a;q)_n &= \prod_{k=0}^{n-1}(1-aq^k),\\
(a)_{-n}:=(a;q)_{-n} &= \frac{1}{\prod_{k=1}^{n}(1-aq^{-k})}.
\end{split}
\end{equation*}
Following~\cite{BMS:1995}, we are going to use an extended definition in this 
paper called the bilateral Bailey pair.
A pair $(\alpha_n,\beta_n)$ of sequences $\{\alpha_n\}_{n\in\Z}$ and
$\{\beta_n\}_{n\in\Z}$ is said to be a \textbf{bilateral Bailey pair} with respect 
to $a$ if 
\begin{equation}\label{eq:defbail}
\beta_n=\sum_{j=-\infty}^n \frac{\alpha_j}{(q)_{n-j}(aq)_{n+j}}.
\end{equation}

\begin{theorem}[\textbf{Bilateral Bailey lemma}~\cite{A:1984,B:1949,BMS:1995}]
If $(\alpha_n,\beta_n)$ is a bilateral Bailey pair then 
\begin{equation}\label{eq:bilaterlemma}
\begin{split}
\sum_{n=-\infty}^{\infty}(\rho_1)_n(\rho_2)_n & (aq/\rho_1\rho_2)^n\beta_n\\
= & \frac{(aq/\rho_1)_{\infty}(aq/\rho_2)_{\infty}}{(aq)_{\infty}(aq/\rho_1 
\rho_2)_{\infty}}\sum_{n=-\infty}^{\infty} \frac{(\rho_1)_n(\rho_2)_n
(aq/\rho_1\rho_2)^n\alpha_n}{(aq/\rho_1)_n(aq/\rho_2)_n}.
\end{split}
\end{equation}
\end{theorem}

This lemma has been used with various Bailey pairs and different specializations 
of the parameters $\rho_1$ and $\rho_2$ to prove many $q$-series identities 
(see for example~\cite{AAB:1987,BMS:1995,FQ:1996,S:1951}).
In this paper the bilateral Bailey lemma is used to derive character identities
for $N=1,2$ superconformal algebras from nonunitary minimal models 
$M(p,p')$.

A useful way to obtain new Bailey pairs from old ones is the construction of
dual Bailey pairs. If $(\alpha_n,\beta_n)$ is a bilateral Bailey pair 
with respect to $a$, the \textbf{dual Bailey pair} $(A_n,B_n)$ is defined as  
\begin{equation}\label{eq:dualpair}
\begin{split}
A_n(a,q) &= a^nq^{n^2}\alpha_n(a^{-1},q^{-1}),\\
B_n(a,q) &= a^{-n}q^{-n^2-n}\beta_n(a^{-1},q^{-1}).
\end{split}
\end{equation} 
Then $(A_n,B_n)$ satisfies \eqref{eq:defbail} with respect to $a$.

\subsection{Bailey pairs from the minimal models $M(p,p')$}
As shown by Foda and Quano~\cite{FQ:1996}, the Bose-Fermi character 
identities~\cite{BM:1994,BMS:1997,FW:2001,W:2002} for the minimal models $M(p,p')$
are of the form 
\begin{equation}\label{eq:bosefermi1}
B_{r(b),s}(L,b;q)=q^{-\mathcal{N}_{r(b),s}}F_{r(b),s}(L,b;q),
\end{equation}
with $\mathcal{N}_{r(b),s}$ as given in~\cite{BMS:1997} and 
\begin{equation}\label{eq:brs1}
\begin{split}
B_{r(b),s}(L,b;q)
=\sum_{j=-\infty}^{\infty} & \Bigg( q^{j(jpp'+r(b)p'-sp)}\left[\begin{array}{c}
L\\ \frac{1}{2}(L+s-b)-jp' \end{array}\right]_q \\ 
&-q^{(jp-r)(jp'-s)}\left[\begin{array}{c}L\\ \frac{1}{2}(L-s-b)+jp' \end{array}\right]_q 
\Bigg).
\end{split}
\end{equation}
Here 
\begin{equation}
\left[\begin{array}{c}n\\ j \end{array}\right]_q=\frac{(q)_n}{(q)_j(q)_{n-j}}
\end{equation}
is the $q$-binomial coefficient. The function fermionic formula $F_{r(b),s}(L,b;q)$
will be discussed in the next section. For simplicity we are going to write $r$ 
for $r(b)$. Following~\cite{FQ:1996,BMS:1995} the identity~\eqref{eq:bosefermi1}
yields the bilateral Bailey pair relative to $a=q^{b-s+2x}$ where $x=\frac{L-2n-b+s}{2}$
\begin{equation}\label{eq:bail1-1}
\begin{split}
\alpha_n &= \begin{cases}
q^{j(jpp'+rp'-sp)} & \text{if $n=jp'-x$}\\
-q^{(jp-r)(jp'-s)} & \text{if $n=jp'-b-x$}\\
0 & \text{otherwise}
\end{cases}\\
\beta_n &= \frac{q^{-\mathcal{N}_{r,s}}}{\left(aq\right)_{2n}} 
F_{r,s}^{(p,p')}(2n+b-s+2x,b;q).
\end{split}
\end{equation}
The dual Bailey pair to \eqref{eq:bail1-1} relative to $a=q^{b-s+2x}$ is
\begin{equation}\label{eq:dbail1-1}
\begin{split}
\hat{\alpha}_n &= \begin{cases}
q^{j^2p'(p'-p)-jp'(r-b)-js(p'-p)-x(b+x-s)} & \text{if $n=jp'-x$}\\
-q^{(jp'-s)(j(p'-p)+r-b)-x(b+x-s)} & \text{if $n=jp'-b-x$}\\
0 & \text{otherwise}
\end{cases}\\
\hat{\beta}_n &= \frac{q^{\mathcal{N}_{r,s}}}{\left(aq\right)_{2n}} 
a^n q^{n^2}F_{r,s}^{(p,p')}(2n+b-s+2x,b;q^{-1}).
\end{split}
\end{equation}
Inserting \eqref{eq:bail1-1} and \eqref{eq:dbail1-1} into the bilateral Bailey 
lemma yields
\begin{equation}\label{eq:ra1-1}
\begin{split}
&\sum_{n=0}^{\infty}(\rho_1)_n(\rho_2)_n(aq/\rho_1\rho_2)^n\frac{q^{-\mathcal{N}_
{r(b),s}}}{\left(aq\right)_{2n}} F_{r,s}^{(p,p')}(2n+b-s+2x,b;q)\\
&=\frac{(aq/\rho_1)_{\infty}(aq/\rho_2)_{\infty}}{(aq)_{\infty}
(aq/\rho_1 \rho_2)_{\infty}}
\sum_{j=-\infty}^{\infty} \Bigg(\frac{(\rho_1)_{jp'-x}(\rho_2)_{jp'-x}}{(aq/\rho_1)_
{jp'-x}(aq/\rho_2)_{jp'-x}}(aq/\rho_1\rho_2)^{jp'-x}\\ & \times q^{j(jpp'+rp'-sp)}
-\frac{(\rho_1)_{jp'-b-x}(\rho_2)_{jp'-b-x}}{(aq/\rho_1)_
{jp'-b-x}(aq/\rho_2)_{jp'-b-x}}\\ & \times (aq/\rho_1\rho_2)^{jp'-b-x} q^{(jp-r)(jp'-s)}
\Bigg)
\end{split}
\end{equation}
and
\begin{equation}\label{eq:ra1-2}
\begin{split}
&\sum_{n=0}^{\infty}(\rho_1)_n(\rho_2)_n(aq/\rho_1\rho_2)^n\frac{q^{\mathcal{N}_
{r(b),s}}}{\left(aq\right)_{2n}}a^n q^{n^2} F_{r,s}^{(p,p')}(2n+b-s+2x,b;q^{-1})\\
&=\frac{(aq/\rho_1)_{\infty}(aq/\rho_2)_{\infty}}{(aq)_{\infty}
(aq/\rho_1 \rho_2)_{\infty}}
\sum_{j=-\infty}^{\infty} \Bigg(\frac{(\rho_1)_{jp'-x}(\rho_2)_{jp'-x}}{(aq/\rho_1)_
{jp'-x}(aq/\rho_2)_{jp'-x}}(aq/\rho_1\rho_2)^{jp'-x}\\ & \times 
q^{j^2p'(p'-p)-jp'(r-b)-js(p'-p)-x(b+x-s)}
-\frac{(\rho_1)_{jp'-b-x}(\rho_2)_{jp'-b-x}}{(aq/\rho_1)_
{jp'-b-x}(aq/\rho_2)_{jp'-b-x}}\\ & \times (aq/\rho_1\rho_2)^{jp'-b-x} 
q^{(jp'-s)(j(p'-p)+r-b)-x(b+x-s)}
\Bigg).
\end{split}
\end{equation}
As in~\cite{BMS:1995}, we are going to consider different specializations of the 
parameters $\rho_1$ and $\rho_2$ in \eqref{eq:ra1-1} and \eqref{eq:ra1-2} to get
character identities for $N=1,2$ superconformal algebras.

\subsection{Fermionic formulas for $M(p,p')$} 
So far we have only considered the bosonic side of \eqref{eq:bosefermi1} explicitly. 
It suffices for the purpose of this paper to state the fermionic formula for the 
case $p<p'<2p$ with $p$ and $p'$ relatively prime and  $r,s$ being pure Takahashi 
length. We follow~\cite[Section 4]{BMSW:1997}. The fermionic formula depends on 
the continued fraction decomposition
\begin{equation*}
\frac{p'}{p'-p}=1+\nu_0 +\cfrac{1}{\nu_1+ \cfrac{1}{\nu_2 +\cdots 
\cfrac{1}{\nu_{n_0}+2}}}.
\end{equation*}
Define $t_i=\sum_{j=0}^{i-1}\nu_j$ for $1\le i \le n_0+1$ and the fractional
level incidence matrix $\I_B$ and corresponding Cartan matrix $B$ as
\begin{equation*}
\begin{split}
(\I_B)_{j,k}&=\begin{cases}  
     \delta_{j,k+1}+\delta_{j,k-1} & \text{for $1\le j<t_{n_0+1},j\ne t_i$}\\
     \delta_{j,k+1}+\delta_{j,k}-\delta_{j,k-1} & \text{for $j= t_i, 1\le i 
                                               \le n_0-\delta_{\nu_{n_0},0}$}\\
     \delta_{j,k+1}+\delta_{\nu_{n_0},0}\delta_{j,k} & \text{for $j= t_{n_0+1}$}
\end{cases}\\
B&=2I_{t_{n_0+1}}-\I_B,
\end{split}
\end{equation*}
where $I_{n}$ is the identity matrix of dimension $n$. Recursively define
\begin{equation*}
\begin{split}
y_{m+1}&=y_{m-1}+(\nu_m+\delta_{m,0}+2\delta_{m,n_0}) y_m, \qquad y_{-1}=0,
\qquad y_0=1,\\
\yb_{m+1}&=\yb_{m-1}+(\nu_m+\delta_{m,0}+2\delta_{m,n_0}) \yb_m, \qquad \yb_{-1}=-1,
\qquad \yb_0=1.
\end{split}
\end{equation*}
Then the Takahashi length and truncated Takahashi length are given by
\begin{equation*}
\begin{array}{ll}
\ell_{j+1}=y_{m-1}+(j-t_m)y_m\\
\lb_{j+1}=\yb_{m-1}+(j-t_m)\yb_m
\end{array}
\qquad \text{for $t_m<j\le t_{m+1}+\delta_{m,n_0}$ with $0\le m\le n_0$.}
\end{equation*}

For $b=\ell_{\beta+1}$, $r(b)=\lb_{\beta+1}$ with $t_{\xi}<\beta\le t_{\xi+1}
+\delta_{\xi,n_0}$ and $s=\ell_{\sigma+1}$ with $t_{\zeta}<\sigma\le t_{\zeta+1}
+\delta_{\zeta,n_0}$ the fermionic formula is given by
\begin{equation}\label{eq:fermi1-1}
F_{r,s}^{(p,p')}(L,b;q)=q^{k_{b,s}}\sum_{\m\equiv Q_{\uu,\vv}\pmod{2}}
q^{\frac{1}{4}\m^tB\m -\frac{1}{2}
A_{\uu,\vv}\m}\prod_{j=1}^{t_{n_0+1}}\left[ \begin{array}{c} n_j+m_j\\m_j \end{array} 
\right]_{q}^{'}
\end{equation}
where $k_{b,s}$ is a normalization constant and $\n,\m\in\Z^{t_{n_0+1}}$ such that
\begin{equation}\label{I-B}  
\n +\m=\frac{1}{2}\Big(\I_B\m+\uu+\vv+L \e_1\Big)
\end{equation}
with $\e_i$ the standard $i$-th basis element of $\Z^{t_{n_0+1}}$,
$\uu=\e_{\beta}-\sum_{k=\xi+1}^{n_0}\e_{t_k}$, $\vv=\e_\sigma-\sum_{k=\zeta+1}^{n_0}
\e_{t_k}$ and $Q_{\uu,\vv}$, $A_{\uu,\vv}$ as defined in~\cite[Section 4.2]{BMSW:1997}.
The $q$-binomial is also defined for negative entries
\begin{equation*}
\left[ \begin{array}{c} n+m\\m \end{array}\right]_q^{'}=
\frac{(q^{n+1})_m}{(q)_m}.
\end{equation*}
Note that
\begin{equation}\label{eq:inverbinom}
\left[ \begin{array}{c} n+m\\m \end{array} 
\right]_{q^{-1}}^{'} = q^{-nm}\left[ \begin{array}{c} n+m\\m \end{array} 
\right]_q^{'}.
\end{equation}
In fact using \eqref{eq:inverbinom} we get the following dual form of the fermionic 
formula that will be useful later on
\begin{multline}\label{eq:fermi1-2}
F_{r,s}^{(p,p')}(L,b;q^{-1})=\\q^{-k_{b,s}}\sum_{\m\equiv Q_{\uu,\vv}}   
q^{\frac{1}{4}\m^tB\m -\frac{1}{2}Lm_1+\frac{1}{2}A_{\uu,\vv}\m-\frac{1}{2}\m^t(\uu+\vv)}
\prod_{j=1}^{t_{n_0+1}}\left[ 
\begin{array}{c} n_j+m_j\\m_j \end{array} \right]_q^{'}.
\end{multline}

\section{$N=1$ Superconformal character from $M(p,p')$} \label{sec:N1}

In this section we are going to consider the specialization in~\eqref{eq:ra1-1} 
and~\eqref{eq:ra1-2}
\begin{equation}
\rho_1\longrightarrow \infty,\quad \rho_2=\text{finite}.
\end{equation} 
We will see that these give characters of the $N=1$ superconformal model $SM(p,p')$ 
given by~\cite{Dobrev:1987,GKO:1986},
\begin{equation}\label{eq:N1char}
\tilde{\chi}^{(p,p')}_{r,s}(q)=\tilde{\chi}^{(p,p')}_{p-r,p'-s}(q)
=\frac{(-q^{\epsilon_{r-s}})_{\infty}}{(q)_{\infty}}
\sum_{j=-\infty}^{\infty}\Big(q^{\frac{j(jpp'+rp'-sp)}{2}}-q^{\frac{(jp-r)(jp'-s)}{2}}
\Big),
\end{equation}
where $1\le r\le p-1,1\le s\le p'-1$, $p$ and $(p'-p)/2$ are relatively prime and
\begin{equation}
 \epsilon_i=
 \begin{cases} \frac{1}{2}& \text{if $i$ is even (NS-sector),}\\
            1 & \text{ if $i$ is odd (R-sector).}
\end{cases}              
\end{equation}
The central charge is $c=\frac{3}{2}-\frac{3(p-p')^2}{pp'}$.

\subsection{The model $SM(p',2p+p')$}
Specializing $\rho_1\longrightarrow \infty$ and $\rho_2=-q^{\frac{b-s+1}{2}}$ with 
$x=0$ in \eqref{eq:ra1-1} we find for $b-s$ even (NS sector)
\begin{equation}\label{eq:N1NS-1}
\tilde{\chi}^{(p',2p+p')}_{s,2r+b}(q)= \sum_{n\ge 0} \frac{q^{\frac{1}{2}(n^2+nb-ns)}
(-q^{\frac{1}{2}})_{n+(b-s)/2}}{(q)_{2n+b-s}}q^{-\mathcal{N}_{r,s}}F_{r,s}^{(p,p')}(2n+b-s,b;q)
\end{equation}
and for $b-s$ odd (R-sector)
\begin{equation}\label{eq:N1R-1}
\tilde{\chi}^{(p',2p+p')}_{s,2r+b}(q)=\sum_{n\ge 0} \frac{q^{\frac{1}{2}(n^2+nb-ns)}
(-q)_{n+(b-s-1)/2}}{(q)_{2n+b-s}}q^{-\mathcal{N}_{r,s}}F_{r,s}^{(p,p')}(2n+b-s,b;q).
\end{equation}

To obtain an explicit fermionic formula set $m_0=L=2n+b-s$ and insert \eqref{eq:fermi1-1}
into \eqref{eq:N1NS-1}. Then using
\begin{equation}\label{eq:-q1/2}
(-q^{\frac{1}{2}})_{\frac{m_0}{2}}=\sum_{k=0}^{\frac{m_0}{2}}q^{\frac{1}{2}(\frac{m_0}{2}-k)^2}
\left[ \begin{array}{c}\frac{m_0}{2} \\ k \end{array} \right]_q
\end{equation}
we find
\begin{equation}\label{eq:fermiN1NS-1}
\begin{split}
\tilde{\chi}^{(p',2p+p')}_{s,2r+b}(q) & = q^{-\frac{1}{8}(b-s)^2-\mathcal{N}_{r,s}+k_{b,s}}
 \sum_{\substack{m_0=0\\ \text{$m_0$ even}}}^{\infty}\sum_{k=0}^{\frac{m_0}{2}}
 \sum_{\m\equiv Q_{\uu,\vv}}q^{\frac{1}{8}m_0^2+\frac{1}{2}(\frac{m_0}{2}-k)^2}\\ 
& \times q^{\frac{1}{4}\m^tB\m-\frac{1}{2}A_{\uu,\vv}\m} 
 \times \frac{1}{(q)_{m_0}} \left[ \begin{array}{c}\frac{m_0}{2} \\ k \end{array} \right]_q
 \prod_{j=1}^{t_{n_0+1}}\left[ 
\begin{array}{c} n_j+m_j\\m_j \end{array} \right]_q^{'}.
\end{split}
\end{equation}

Setting $\p=(k,m_0,\m) \in \Z^{t_{n_0+1}+2}$, \eqref{eq:fermiN1NS-1} in the NS-sector
can be rewritten as  
\begin{equation}\label{eq:fermiN1NS-2}
\begin{split}
\tilde{\chi}^{(p',2p+p')}_{s,2r+b}(q) &=q^{-\frac{1}{8}(b-s)^2-\mathcal{N}_{r,s}+k_{b,s}}
 \sum_{\substack{\p\in\Z^{t_{n_0+1}+2}\\p_i \equiv (\tilde{Q}_{\uu,\vv})_i,i\ge 2}}
 q^{\frac{1}{4}\p^t\tilde{B}\p-\frac{1}{2}\tilde{A}_{\uu,\vv}\p}\\
 & \times \frac{1}{(q)_{p_2}} \prod_{\substack{j=1,j\ne 2}}^{t_{n_0+1}+2}\left[ 
\begin{array}{c} \frac{1}{2}(\I_{\tilde{B}}\p+\tilde{\uu}+\tilde{\vv})_j \\ p_j \end{array} 
\right]_q^{'}
\end{split}
\end{equation}
where $\I_{\tilde{B}}=2I_{t_{n_0+1}+2}-\tilde{B}$,
\begin{equation}\label{eq:C-1}
\begin{split}
\tilde{B} &= \left( \begin{array}{cc|cc}
                            2  & -1 & 0\\ 
			    -1 & 1 & 1\\\hline  
			    0 & -1 & B \\
			    \end{array}
         \right)\\
\tilde{A}_{\uu,\vv}&=(0,0,A_{\uu,\vv}),\\
\tilde{\uu}^t&=(0,0,\uu^t),\\
\tilde{\vv}^t&=(0,0,\vv^t),\\
\tilde{Q}_{\uu,\vv}^t&=(0,0,Q_{\uu,\vv}^t).
\end{split}
\end{equation}

Similarly setting $m_0=2n+b-s$ in (\ref{eq:N1R-1}) and using 
\begin{equation}\label{eq:-q}
(-q)_{\frac{m_0-1}{2}}=\frac{1}{2}\sum_{k=0}^{\frac{m_0+1}{2}}q^{\frac{1}{2}(\frac{m_0+1}{2}-k)
(\frac{m_0-1}{2}-k)}
\left[ \begin{array}{c}\frac{m_0+1}{2} \\ k \end{array} \right]_q
\end{equation}

we get the fermionic formula in the R-sector,
\begin{multline}\label{eq:fermiN1R-1}
\tilde{\chi}^{(p',2p+p')}_{s,2r+b}(q)  =\frac{1}{2}
 q^{-\frac{1}{8}((b-s)^2+1)-\mathcal{N}_{r,s}+k_{b,s}}
 \sum_{\substack{\p \in \Z^{t_{n_0+1}+2}\\ p_i\equiv (\tilde{Q}_{\uu,\vv})_i, i\ge 2}}
 q^{\frac{1}{4}\p^t\tilde{B}\p-\frac{1}{2}\tilde{A}_{\uu,\vv}\p}\\
  \times \frac{1}{(q)_{p_2}}  \prod_{\substack{j=1,j\ne 2}}^{t_{n_0+1}+2}\left[ 
\begin{array}{c} \frac{1}{2}(\I_{\tilde{B}}\p+\tilde{\uu}+\tilde{\vv})_j \\ p_j \end{array} 
\right]_q^{'}
\end{multline}
where $\tilde{B},\tilde{A},\tilde{\vv}$ are as in \eqref{eq:C-1} and 
$\tilde{\uu}^t=(1,0,\uu^t)$, $\tilde{Q}_{\uu,\vv}^t=(0,1,Q_{\uu,\vv}^t)$.

\subsection{The model $SM(p',3p'-2p)$} Similarly using the same specialization with the dual 
Bailey pair in \eqref{eq:ra1-2} we find for $b-s$ even in the NS-sector
\begin{equation}\label{eq:N1NS-2}
\tilde{\chi}^{(p',3p'-2p)}_{s,3b-2r}(q)=
\sum_{n\ge 0} \frac{q^{\frac{3n}{2}(n+b-s)}(-q^{\frac{1}{2}})_{n+(b-s)/2}}{(q)_{2n+b-s}}  
              q^{\mathcal{N}_{r,s}}F_{r,s}^{(p,p')}(2n+b-s,b;q^{-1}) 
\end{equation}
and for $b-s$ odd in the R-sector
\begin{equation}\label{eq:N1R-2}
\tilde{\chi}^{(p',3p'-2p)}_{s,3b-2r}(q)=
\sum_{n\ge 0} \frac{q^{\frac{3n}{2}(n+b-s)}(-q)_{n+(b-s-1)/2}}{(q)_{2n+b-s}}  
              q^{\mathcal{N}_{r,s}}F_{r,s}^{(p,p')}(2n+b-s,b;q^{-1}).
\end{equation}               

To obtain the fermionic formula, as before we are going to set $m_0=2n+b-s$.
Inserting \eqref{eq:-q} and \eqref{eq:fermi1-2} into \eqref{eq:N1R-2}
we get in the R-sector
\begin{equation}\label{eq:fermiN1R-2}
\begin{split}
\tilde{\chi}^{(p',3p'-2p)}_{s,3b-2r}(q) &= \frac{1}{2}q^{-\frac{1}{8}(3(b-s)^2+1)+ 
\mathcal{N}_{r,s}-k_{b,s}}
 \sum_{\substack{m_0=0\\ \text{$m_0$ odd}}}^{\infty}\sum_{k=0}
^{\frac{m_0+1}{2}}\sum_{\m\equiv Q_{\uu,\vv}}\\
&\times q^{\frac{1}{2}(m_0^2+k^2-m_0k-m_0m_1)} 
  q^{\frac{1}{4}\m^tB\m-\frac{1}{2}\m^t(\uu+\vv)+\frac{1}{2}A_{\uu,\vv}\m}\\ 
& \times \frac{1}{(q)_{m_0}} \left[ \begin{array}{c}\frac{m_0+1}{2} \\ k \end{array} \right]_q
 \prod_{j=1}^{t_{n_0+1}}\left[ 
\begin{array}{c} n_j+m_j\\m_j \end{array} \right]_q^{'}.
\end{split}
\end{equation}
Define $\p=(k,m_0,\m) \in \Z^{t_{n_0+1}+2}$, so that \eqref{eq:fermiN1R-2} in the R-sector
can be rewritten as
\begin{multline}\label{eq:fermiN1R-3}
\tilde{\chi}^{(p',3p'-2p)}_{s,3b-2r}(q) =\frac{1}{2}
 q^{-\frac{1}{8}(3(b-s)^2+1)+\mathcal{N}_{r,s}-k_{b,s}}
 \sum_{\substack{\p\in \Z^{t_{n_0+1}+2}\\ p_i\equiv (\tilde{Q'}_{\uu,\vv})_i, i\ge 2}}
q^{\frac{1}{4}\p^t\tilde{B}'\p+\frac{1}{2}\tilde{A}_{\uu,\vv}\p}\\
 \times \frac{1}{(q)_{p_2}} \prod_{\substack{j=1,j\ne 2}}^{t_{n_0+1}+2}\left[ 
\begin{array}{c} \frac{1}{2}(\I_{\tilde{B}'}\p+\tilde{\uu}+\tilde{\vv})_j \\ p_j \end{array} 
\right]_q^{'}
\end{multline}
where $\I_{\tilde{B}'}=2I_{t_{n_0+1}+2}-\tilde{B}'$, $\tilde{\vv}$ as in \eqref{eq:C-1}, 
$\tilde{\uu}^t =(1,0,\uu^t)$, $(\tilde{Q}_{\uu,\vv}')^t =(0,1,Q_{\uu,\vv}^t)$, and
\begin{equation}\label{eq:C'-1}
\begin{split}
\tilde{B}'&=\left( \begin{array}{cc|cc}
                            2  & -1 & 0\\ 
			    -1 & 2 & -1\\\hline  
			    0 & -1 & B \\
			    \end{array}
         \right)\\
\tilde{A}_{\uu,\vv}&=(0,0,A_{\uu,\vv}-\uu^t-\vv^t).
\end{split}
\end{equation}

Similarly, for the NS-sector it follows from \eqref{eq:N1NS-2}
\begin{multline}\label{eq:fermiN1NS-3}
\tilde{\chi}^{(p',3p'-2p)}_{s,3b-2r}(q) =q^{-\frac{3}{8}(b-s)^2+\mathcal{N}_{r,s}-k_{b,s}}
 \sum_{\substack{\p\in\Z^{t_{n_0+1}+2}\\ p_i\equiv (\tilde{Q'}_{\uu,\vv})_i, i\ge 2}}
 q^{\frac{1}{4}\p^t\tilde{B}'\p+\frac{1}{2}\tilde{A}_{\uu,\vv}\p}\\
  \times \frac{1}{(q)_{p_2}} \prod_{\substack{j=1,j\ne 2}}^{t_{n_0+1}+2}\left[ 
\begin{array}{c} \frac{1}{2}(\I_{\tilde{B}'}\p+\tilde{\uu}+\tilde{\vv})_j \\ p_j \end{array} 
\right]_q^{'}
\end{multline}
with $\tilde{B}'$ and $\tilde{A}_{\uu,\vv}$ as in \eqref{eq:C'-1},
$(\tilde{Q}_{\uu,\vv}')^t=(0,0,Q_{\uu,\vv}^t)$, $\tilde{\uu}^t =(0,0,\uu^t)$
and $\tilde{\vv}^t =(0,0,\vv^t)$.

\section{$N=2$ Character formulas} \label{sec:N2}
\subsection{$N=2$ superconformal algebra and Spectral flow}
The $N=2$ superconformal algebra $\mathcal{A}$ is the infinite dimensional Lie 
super algebra~\cite{EG:1996} with basis $L_n,T_n,G_r^{\pm},C$ and (anti)-commutation 
relation given by
\begin{equation*}
\begin{split}
\left[L_m,L_n\right] &= (m-n)L_{m+n}+\frac{C}{12}(m^3-m)\delta_{m+n,0}\\
\left[L_m,G_r^{\pm}\right] &= (\frac{1}{2}m-r)G_{m+r}^{\pm}\\
\left[L_m,T_n\right]&= -nT_{m+n}\\
\left[T_m,T_n\right] &= \frac{1}{3}cm\delta_{m+n,0}\\
\left[T_m,G_r^{\pm}\right] &= \pm G_{m+r}^{\pm}\\
\{G_r^+,G_s^-\} &= 2L_{r+s}+(r-s)T_{r+s}+\frac{C}{3}(r^2-\frac{1}{4})
\delta_{r+s,0}\\
\left[L_m,C\right]&=\left[T_n,C\right]=\left[G_r^{\pm},C\right]=0\\
\{ G_r^+,G_s^+\} &= \{ G_r^-,G_s^-\}=0
\end{split}
\end{equation*}
where $n,m\in\Z$, but $r,s$ are integers in R-sector and half-integer in NS-sector.
The element $C$ is the central element and its eigenvalue $c$ is parametrized as 
$c=3(1-\frac{2p}{p'})$, where $p,p'$ are relatively prime positive integers.

It was observed in~\cite{K:2003,SS:1987} that there exits a family of outer 
automorphisms $\alpha_{\eta}:\mathcal{A}\rightarrow \mathcal{A}$ which maps the 
$N=2$ superconformal algebras to itself. These are explicitly given by
\begin{equation}\label{eq:auto}
\begin{split}
\alpha_{\eta}(G_r^+)&=\hat{G}_r^+=G_{r-\eta}^+\\
\alpha_{\eta}(G_r^-)&=\hat{G}_r^-=G_{r+\eta}^-\\
\alpha_{\eta}(L_n)&=\hat{L}_n=L_n-\eta T_n +\frac{c}{6}\eta^2\delta_{n,0}\\
\alpha_{\eta}(T_n)&=\hat{T}_n=T_n-\frac{c}{3}\eta \delta_{n,0}
\end{split}
\end{equation}
This family of automorphisms is called \textbf{spectral flow} and $\eta \in \R$ is 
called the \textbf{flow parameter}. 
When $\eta \in \Z$ each sector of the algebra is mapped to itself.
When $\eta \in \Z+\frac{1}{2}$ the Neveu-Schwarz sector is mapped to the Ramond sector 
and vice-versa. We are going to use the spectral flow $\eta=\pm \frac{1}{2}$ to map 
the NS-sector to the R-sector.

\subsection{Spectral flow and characters}
We denote the Verma module generated from a highest weight state $|h,Q,c\rangle$ with 
$L_0$ eigenvalue $h$, $T_0$ eigenvalue $Q$ and central charge $c$ by $V_{h,Q}$. 
The character $\chi_{V_{h,Q}}$ of a highest weight representation $V_{h,Q}$ is 
defined as
\begin{equation*}
\chi_{V_{h,Q}}(q,z)=\mathrm{Tr}_{V_{h,Q}}(q^{L_0-c/24}z^{T_0}).
\end{equation*}
Following~\cite{K:2003} the character transforms under the spectral flow in the 
following way 
\begin{equation}\label{eq:specchar}
\mathrm{Tr}_{V_{h,Q}}(q^{\hat{L}_0-c/24}z^{\hat{T}_0})
=\mathrm{Tr}_{V_{h^{\eta},Q^{\eta}}}(q^{L_0-c/24}z^{T_0}),
\end{equation}
where $h^{\eta}$ and $Q^{\eta}$ are the eigenvalues of $\hat{L}_0$ and $\hat{T}_0$, 
respectively, as defined in \eqref{eq:auto}. This means the new character 
$\chi_{V_{h^{\eta},Q^{\eta}}}(q,z)$ which is the trace of the transformed operators
over the original representation equals the character of the representation defined 
by the eigenvalues $h^{\eta}$ and $Q^{\eta}$ of $\hat{L}_0$ and $\hat{T}_0$, 
respectively. So the new character is the character of the representation 
$V_{h^{\eta},Q^{\eta}}$.

For $\eta=\frac{1}{2}$ the spectral flow $\alpha_{\frac{1}{2}}$ takes a
NS-sector character to an R-sector character. Let $\chi_{V_{h,Q}}^{NS}(q,z)$ be a 
NS-sector character corresponding to the representation  $V_{h,Q}$. Then 
by~\eqref{eq:specchar} and~\eqref{eq:auto} the new R-sector character
$\chi_{V_{h^{\eta},Q^{\eta}}}^{R}(q,z)$ is derived using
\begin{equation}\label{eq:specns}
\begin{split}
\chi_{V_{h^{\eta},Q^{\eta}}}^{R}(q,z)
&=\mathrm{Tr}_{V_{h,Q}}(q^{\hat{L}_0-c/24}z^{\hat{T}_0})
=\mathrm{Tr}_{V_{h,Q}}(q^{L_0-\frac{1}{2}T_0+\frac{c}{24}-\frac{c}{24}}
z^{T_0-\frac{c}{6}})\\
&=q^{\frac{c}{24}}z^{-\frac{c}{6}}\mathrm{Tr}_{V_{h,Q}}(q^{L_0-\frac{c}{24}}{
(zq^{-\frac{1}{2}})}^{T_0})
=q^{\frac{c}{24}}z^{-\frac{c}{6}}\chi_{V_{h,Q}}^{NS}(q,zq^{-\frac{1}{2}}).
\end{split}
\end{equation}

\subsection{R-sector character from NS-sector character} 
{}To simplify notation we are going to use a slightly different notation for characters. 
Since we are only dealing with the vacuum character in the NS-sector for which $h=0,Q=0$,
we write $\hat{\chi}^{NS}_{p,p'}(q,z)$. The R-sector character is denoted by
$\hat{\chi}^{R}_{p,p'}(q,z)$ with the corresponding $(h,Q)$ specified separately.

Following~\cite{Dobrev:1986,D:1998,EG:1996,K:2003,K:2004} the vacuum character for 
the $N=2$ superconformal algebra with central element $c=3(1-\frac{2p}{p'})$ in the 
NS-sector is given by
\begin{multline}\label{eq:vacuum1}
\hat{\chi}_{p,p'}^{NS}(q,z)=q^{-c/24}\prod_{n=1}^{\infty}\frac{(1+zq^{n-
\frac{1}{2}})(1+z^{-1}q^{n-\frac{1}{2}})}{{(1-q^n)}^2}\\
 \times \Big(1-\sum_{n=0}^\infty \bigl(q^{p(n+1)(p'(n+1)-1)}
 +\frac{zq^{p'n(pn+1)+pn+\frac{1}{2}}}{1+zq^{p'n+\frac{1}{2}}}+
 \frac{z^{-1}q^{p'n(pn+1)+pn+\frac{1}{2}}}{1+z^{-1}q^{p'n+\frac{1}{2}}}\bigr)\\
 +\sum_{n=1}^\infty \bigl(q^{pn(p'n+1)}+\frac{zq^{p'n(pn+1)-pn-
\frac{1}{2}}}{1+zq^{p'n-\frac{1}{2}}}+\frac{z^{-1}q^{p'n(pn+1)-pn-
\frac{1}{2}}}{1+z^{-1}q^{p'n-\frac{1}{2}}}\bigr)\Big).
\end{multline}
This formula can be verified using the embedding diagram for the vacuum character as 
described in~\cite{EG:1996,K:2003} and can be rewritten as (as will be useful later)
\begin{multline}\label{eq:vacuum2}
\hat{\chi}^{NS}_{p,p'}(q,z)=q^{-c/24}\prod_{n=1}^{\infty} 
\frac{(1+zq^{n-\frac{1}{2}})(1+z^{-1}q^{n-\frac{1}{2}})}{(1-q^n)^2}\\
\times \sum_{j=-\infty}^{\infty}q^{pj(p'j+1)}\frac{1-q^{2p'j+1}}
{(1+zq^{p'j+\frac{1}{2}})(1+z^{-1}q^{p'j+\frac{1}{2}})}.
\end{multline}
The unitary case $p=1$ of these character formulas was given 
in~\cite{Dobrev:1987a,Kirt:1988,M:1987,RY:1987}.
In particular if we put $z=1$ in \eqref{eq:vacuum2} we obtain the following formula 
derived in~\cite{EG:1996}
\begin{equation}\label{eq:vacuum3}
\hat{\chi}^{NS}_{p,p'}(q)=q^{-c/24}\prod_{n=1}^{\infty} 
\frac{(1+q^{n-\frac{1}{2}})^2}{(1-q^n)^2}
\sum_{j=-\infty}^{\infty}q^{pj(p'j+1)}\frac{1-q^{p'j+\frac{1}{2}}}
{1+q^{p'j+\frac{1}{2}}}.
\end{equation}

Let us apply \eqref{eq:specns} to the NS-sector vacuum character \eqref{eq:vacuum2}
to get a Ramond sector character. From \eqref{eq:auto} it follows that
\begin{equation*}
\begin{split}
\hat{L}_0&=L_0-\frac{1}{2}T_0+\frac{c}{24}\\
\hat{T}_0&=T_0-\frac{c}{6}.
\end{split}
\end{equation*}

For the vacuum character in the NS-sector $(h,Q)=(0,0)$, so the new eigenvalues 
are $(h^{\eta},Q^{\eta})= (\frac{c}{24},-\frac{c}{6})$ in the R-sector. 
Hence the new character in the R-sector corresponds to $(h^{\eta},Q^{\eta})$ and 
by~\eqref{eq:specns}
\begin{multline}\label{eq:Ramond1}
\hat{\chi}^{R}_{p,p'}(q,z)=q^{\frac{c}{24}}z^{-\frac{c}{6}}
\hat{\chi}^{NS}_{p,p'}(q,zq^{-\frac{1}{2}})\\
=z^{-\frac{c}{6}}\frac{(-z)_{\infty}(-z^{-1}q)_{\infty}}{(q)^2_{\infty}}
\sum_{j=-\infty}^{\infty}q^{pj(p'j+1)}
\frac{1-q^{2p'j+1}}{(1+zq^{p'j})(1+z^{-1}q^{p'j+1})}.
\end{multline}

\subsection{$N=2$ superconformal characters for $c=3(1-\frac{2p}{p'})$}
Using $r=0$ and $b=1$ in \eqref{eq:ra1-1} we obtain
\begin{multline}\label{eq:ra3-1}
\sum_{n=0}^{\infty}(\rho_1)_n(\rho_2)_n(aq/\rho_1\rho_2)^n
\frac{q^{-\mathcal{N}_{0,s}}}{\left(aq\right)_{2n}}
F_{0,s}^{(p,p')}(2n+1-s+2x,1;q)\\
=\frac{(aq/\rho_1)_{\infty}(aq/\rho_2)_{\infty}}{(aq)_{\infty}(aq/\rho_1 \rho_2)_{\infty}}
\sum_{j=-\infty}^{\infty} \Bigg(\frac{(\rho_1)_{jp'-x}(\rho_2)_{jp'-x}}{(aq/\rho_1)_
{jp'-x}(aq/\rho_2)_{jp'-x}}(aq/\rho_1 \rho_2)^{jp'-x}\\
-\frac{(\rho_1)_{jp'-1-x}(\rho_2)_{jp'-1-x}}{(aq/\rho_1)_{jp'-1-x}(aq/\rho_2)_{jp'-1-x}}
(aq/\rho_1 \rho_2)^{jp'-1-x} \Bigg)q^{jp(jp'-s)}.
\end{multline}
In this section we consider the specialization 
\begin{equation*}
\rho_1=\text{finite},\quad \rho_2=\text{finite}.
\end{equation*} 
Taking the limit $\frac{aq}{\rho_1\rho_2}\longrightarrow 1$ in \eqref{eq:ra3-1}, we find
\begin{multline}\label{eq:ra3-2}
\sum_{n=0}^{\infty}(\rho_1)_n(\rho_2)_n\frac{q^{-\mathcal{N}_{0,s}}}
{\left(aq\right)_{2n}}F_{0,s}^{(p,p')}(2n+1-s+2x,1;q)\\
=\frac{(\rho_1)_{\infty}(\rho_2)_{\infty}}{(\rho_1\rho_2)_{\infty}(q)_{\infty}} 
\sum_{j=-\infty}^{\infty}q^{jp(jp'-s)}
 \frac{\rho_1\rho_2q^{2(jp'-x-1)}-1}{(1-\rho_1q^{jp'-x-1})(1-\rho_2q^{jp'-x-1})}.
\end{multline}

\subsection{NS-sector characters}
Let us set $\rho_1=-zq^{x+\frac{1}{2}},\rho_2=-z^{-1}q^{x+\frac{1}{2}}$ 
in~\eqref{eq:ra3-2}, which implies $a=q^{2x}$ and $s=1$. Making the variable
change $j\longrightarrow -j$ in \eqref{eq:ra3-2} and setting $x=0$ we obtain
\begin{multline}\label{eq:ra3-3}
\sum_{n=0}^{\infty}(-zq^{\frac{1}{2}})_n(-z^{-1}q^{\frac{1}{2}})_n
\frac{q^{-\mathcal{N}_{0,1}}}{\left(q\right)_{2n}} F_{0,1}^{(p,p')}(2n,1;q)\\
=\frac{(-zq^{\frac{1}{2}})_{\infty}(-z^{-1}q^{\frac{1}{2}})_{\infty}}
{(q)_{\infty}^2} \sum_{j=-\infty}^{\infty}q^{jp(jp'+1)}
 \frac{1-q^{2jp'+1}}{(1+zq^{jp'+\frac{1}{2}})(1+z^{-1}q^{jp'+\frac{1}{2}})}.
\end{multline}
Comparing with \eqref{eq:vacuum2}, we obtain
\begin{equation}\label{eq:ra3-5}
\hat{\chi}^{NS}_{p,p'}(q,z)=q^{-\frac{c}{24}-\mathcal{N}_{0,1}}
\sum_{n=0}^{\infty} \frac{(-zq^{\frac{1}{2}})_n(-z^{-1}q^{\frac{1}{2}})_n}{\left(q\right)_{2n}}
F_{0,1}^{(p,p')}(2n,1;q).
\end{equation}

Setting $z=1$ and inserting the fermionic formula \eqref{eq:fermi1-1}, we find
\begin{multline}\label{eq:fermi3-1}
\hat{\chi}^{NS}_{p,p'}(q)=q^{-\frac{c}{24}-\mathcal{N}_{0,1}+k_{1,1}}
\sum_{n=0}^{\infty}\Bigg( \frac{{(-q^{\frac{1}{2}})}^2_n}{(q)_{2n}}
 \sum_{\m\equiv Q_{\uu,\vv}} q^{\frac{1}{4}\m^tB\m-\frac{1}{2}A_{\uu,\vv}\m}\\
 \times \prod_{j=1}^{t_{n_0+1}}\left[ \begin{array}{c} n_j+m_j\\m_j \end{array} 
 \right]_q^{'}\Bigg).
\end{multline}
Let us set $m_0=2n$ and use \eqref{eq:-q1/2} to get
\begin{multline}\label{eq:fermi3-2}
\hat{\chi}^{NS}_{p,p'}(q) = q^{-\frac{c}{24}-\mathcal{N}_{0,1}+k_{1,1}} 
\sum_{\substack{m_0=0\\ \text{$m_0$ even}}}^{\infty}\sum_{k_1=0}^{\frac{m_0}{2}}
 \sum_{k_2=0}^{\frac{m_0}{2}}\sum_{\m\equiv Q_{\uu,\vv}}
 q^{\frac{1}{2}(\frac{m_0}{2}-k_1)^2+\frac{1}{2}(\frac{m_0}{2}-k_2)^2}\\
 \times q^{\frac{1}{4}\m^tB\m-\frac{1}{2}A_{\uu,\vv}\m}
 \frac{1}{(q)_{m_0}} \left[ \begin{array}{c}\frac{m_0}{2} \\ k_1 \end{array} 
 \right]_q
\left[ \begin{array}{c}\frac{m_0}{2} \\ k_2 \end{array} \right]_q 
\prod_{j=1}^{t_{n_0+1}}\left[ 
\begin{array}{c} n_j+m_j\\m_j \end{array} \right]_q^{'}.
\end{multline}

Define $\p=(k_1,k_2,m_0,\m)\in \Z^{t_{n_0+1}+3}$, so that \eqref{eq:fermi3-2}
can be rewritten as
\begin{multline}\label{eq:fermi3-3}
\hat{\chi}^{NS}_{p,p'}(q)=q^{-\frac{c}{24}-\mathcal{N}_{0,1}+k_{1,1}}
\sum_{\substack{\p\in\Z^{t_{n_0+1}+3}\\ p_i\equiv (\hat{Q}_{\uu,\vv})_i, i\ge 3}}
q^{\frac{1}{4}\p^tD\p -\frac{1}{2}\hat{A}_{\uu,\vv}\p}\\
 \times \frac{1}{(q)_{p_3}}  \prod_{\substack{j=1,j\ne3}}^{t_{n_0+1}+3}\left[ 
\begin{array}{c} \frac{1}{2}(\I_D\p+\hat{\uu}+\hat{\vv})_j\\p_j \end{array} \right]_q^{'},
\end{multline}
where $\I_D=2I_{t_{n_0+1}+3}-D$ and
\begin{equation}\label{I-D}
\begin{split}
D&=\left( \begin{array}{ccc|cc}
            2 & 0 & -1 & 0\\ 
	    0 & 2 & -1 & 0\\ 
	   -1 & -1 & 1 & 1\\\hline 
            0 & 0 &  -1 & B\\
		\end{array}
         \right),\\
\hat{A}_{\uu,\vv}&=(0,0,0,A_{\uu,\vv}),\\
\hat{\uu}^t&=(0,0,0,\uu^t),\\
\hat{\vv}^t&=(0,0,0,\vv^t),\\
\hat{Q}^t_{\uu,\vv}&=(0,0,0,Q_{\uu,\vv}^t).
\end{split}
\end{equation}
This gives a new fermionic expression for the NS-sector character.

\subsection{Ramond sector characters}
Let us set $\rho_1=-zq^x,\rho_2=-z^{-1}q^{x+1}$ in \eqref{eq:ra3-2}, which implies 
$a=q^{2x}$ and $s=1$. Setting $x=0$ and changing $j\longrightarrow -j$ we obtain
\begin{multline}\label{eq:R2}
\sum_{n=0}^{\infty} \frac{{(-z)}_n{(-z^{-1}q)}_n}{(q)_{2n}} q^{-\mathcal{N}_{0,1}} 
F_{0,1}^{(p,p')}(2n,1;q) \\
=\frac{{(-z)}_{\infty}{(-z^{-1}q)}_{\infty}}{{(q)^2}_{\infty}}\sum_{j=-\infty}^{\infty}
q^{jp(jp'+1)}\frac{1-q^{2jp'+1}}{(1+zq^{jp'})(1+z^{-1}q^{jp'+1})}.
\end{multline}
Comparing with \eqref{eq:Ramond1} we get 
\begin{equation}\label{eq:R3}
\hat{\chi}^{R}_{p,p'}(q,z)=z^{-\frac{c}{6}}q^{-\mathcal{N}_{0,1}}\sum_{n=0}^{\infty} 
\frac{{(-z)}_n{(-z^{-1}q)}_n}{(q)_{2n}}F_{0,1}^{(p,p')}(2n,1;q). 
\end{equation}

Again using \eqref{eq:fermi1-1} in a similar way to the NS-sector and setting $z=1$ we find
\begin{multline}\label{eq:Rfermi3-1}
\hat{\chi}^{R}_{p,p'}(q)= 2q^{-\mathcal{N}_{0,1}+k_{1,1}}
\sum_{n=0}^{\infty}\Bigg( \frac{{(-q)}_{n-1}{(-q)}_n}{(q)_{2n}}
 \sum_{\m\equiv Q_{\uu,\vv}} q^{\frac{1}{4}\m^tB\m-\frac{1}{2}A_{\uu,\vv}\m}\\
 \times \prod_{j=1}^{t_{n_0+1}}\left[ 
\begin{array}{c} n_j+m_j\\m_j \end{array} \right]_q^{'}\Bigg).
\end{multline}
Using
\begin{equation*}
(x)_n =\sum_{k=0}^n (-x)^{(n-k)}q^{\frac{1}{2}(n-k)(n-k-1)}\left[ \begin{array}{c} n\\k \end{array} 
\right]_{q}
\end{equation*}
and setting $m_0=2n$, equation \eqref{eq:Rfermi3-1} can be rewritten as
\begin{multline}\label{eq:Rfermi3-2}
\hat{\chi}^{R}_{p,p'}(q) = 2q^{-\mathcal{N}_{0,1}+k_{1,1}}
\sum_{\substack{m_0=0\\ \text{$m_0$ even}}}^{\infty}\sum_{k_1=0}
^{\frac{m_0}{2}-1}\sum_{k_2=0}^{\frac{m_0}{2}}\sum_{\m\equiv Q_{\uu,\vv}}
 q^{\frac{1}{4}(m_0^2+2k_1^2+2k_2^2-2m_0k_1-2m_0k_2)}\\
 \times q^{\frac{1}{4}\m^tB\m-\frac{1}{2}A_{\uu,\vv}\m+\frac{1}{2}(k_1-k_2)}
 \frac{1}{(q)_{m_0}} \left[ \begin{array}{c}\frac{m_0}{2}-1 \\ k_1 \end{array} \right]_q
\left[ \begin{array}{c}\frac{m_0}{2} \\ k_2 \end{array} \right]_q 
\prod_{j=1}^{t_{n_0+1}}\left[ 
\begin{array}{c} n_j+m_j\\m_j \end{array} \right]_q^{'}.
\end{multline}
Setting $\p=(k_1,k_2,m_0,\m)\in \Z^{t_{n_0+1}+3}$ this becomes
\begin{multline}\label{eq:Rfermi3-3}
\hat{\chi}^{R}_{p,p'}(q)= 2q^{-\mathcal{N}_{0,1}+k_{1,1}}
\sum_{\substack{\p\in \Z^{t_{n_0+1}+3}\\ p_i\equiv (\hat{Q}_{\uu,\vv})_i, i\ge 3}}
q^{\frac{1}{4}\p^tD\p -\frac{1}{2}\hat{A}_{\uu,\vv}\p}\\
 \times \frac{1}{(q)_{p_3}}
\prod_{\substack{j=1,j\ne 3}}^{t_{n_0+1}+3}\left[ 
\begin{array}{c} \frac{1}{2}(\I_D\p+\hat{\uu}+\hat{\vv})_j\\p_j \end{array} \right]_q^{'},
\end{multline}
with the same notations as in \eqref{I-D} except
\begin{equation*}
\begin{split}
\hat{A}_{\uu,\vv}&=(1,-1,0,A_{\uu,\vv}),\quad \hat{\uu}^t=(-1,0,0,\uu^t),\quad
\hat{\vv}^t=(-1,0,0,\vv^t).
\end{split}
\end{equation*} 
This gives a new fermionic expression of the new R-sector character.

\section{Conclusion} \label{sec:conclusion}
In this paper we only considered the vacuum character for the $N=2$ superconformal algebra
with central charge $c=3(1-\frac{2p}{p'})$ with $p<p'$ in the NS-sector and the 
Ramond sector character derived from the vacuum character. We believe that 
similar Bailey flows exist for the general $N=2$ superconformal characters, but explicit
formulas are not yet available in the literature.

The astute reader might have noticed that unlike in section~\ref{sec:N1} we did
not carry out the Bailey flow for the dual Bailey pair in section~\ref{sec:N2}, 
the reason being that the fermionic formula $F^{(p,p')}_{r,s}(L,b;q)$ for $p<p'<2p$ 
and $r=b=1$ are not given in~\cite{BMS:1997,BMSW:1997}. A formula however does appear
in~\cite{W:2002}. The matrix $D$ in this case is
\begin{equation*}
D=\left( \begin{array}{ccc|cc}
            2 & 0 & -1 & 0\\ 
	    0 & 2 & -1 & 0\\ 
	   -1 & -1 & 2 & -1\\\hline 
            0 & 0 &  -1 & B\\
		\end{array}
         \right).
\end{equation*}
Details will be available in~\cite{Deka:2005}.


\end{document}